# Exotic vortex states at high magnetic fields in a quasi-two-dimensional FeSe-based superconductor

Xuyang Li[1,5], Jian Li[1], Kai Liu[1,5], Jiaqiang Cai[2], Shunjiao Li[1], Baolei Kang[1], Mengzhu Shi[1], Dan Zhao[1], Chuanying Xi[2], Jinglei Zhang[2], Tao Wu[1,3,4,5]* and Xianhui Chen[1,3,4,5]*

1. Hefei National Research Center for Physical Sciences at the Microscale, University of Science and Technology of China, Hefei, China

2. Anhui Province Key Laboratory of Condensed Matter Physics at Extreme Conditions, High Magnetic Field Laboratory, Chinese Academy of Sciences, Hefei 230031, China

3. Department of Physics, University of Science and Technology of China, Hefei, Anhui 230026, China

4. Collaborative Innovation Center of Advanced Microstructures, Nanjing University, Nanjing 210093, China

5. Hefei National Laboratory, University of Science and Technology of China, Hefei 230088, China

*Correspondence to: wutao@ustc.edu.cn, chenxh@ustc.edu.cn

**Owing to strong electronic correlations, high-temperature superconductivity always exhibits intricate intertwinement with various competing electronic orders in phase diagrams, such as spin/charge density waves (S/CDWs). In cuprate superconductors, the intertwinement of superconductivity and CDW order could strongly affect the fundamental properties of superconductivity, such as the critical temperature ($T_c$) and critical magnetic field ($H_c$). Recent high-field transport measurements indicate that when quantum fluctuations become important at low temperatures and high magnetic fields, the CDW order also reshapes the vortex states, which leads to fragile superconductivity with extremely low critical current ($J_c$). Here, by performing comprehensive high-field transport measurements, the $H$–$T$ phase diagram of vortex states is mapped to $\mu_0H \sim 33$ T in a quasi-two-dimensional FeSe-based superconductor (TBA$^+$)$_x$FeSe with a zero-resistivity transition temperature above 40 K. Our results indicate that (TBA$^+$)$_x$FeSe is an extremely type II superconductor with significant thermal fluctuations. At low temperatures, high magnetic fields cause the vortex solid state to exhibit similar current-dependent zero-resistance behavior as the fragile superconductivity in cuprate superconductors with CDW order. When the vortex solid state is melted with increasing temperature, a superconducting regime with vortex-like phase fluctuations emerges as an**

**intermediate state, which features finite longitudinal resistance and vanishing Hall resistance. At higher temperatures, a vortex liquid state with finite Hall resistance eventually appears due to thermal fluctuations. All these observations suggest exotic vortex states beyond the classical paradigm of vortex matter. In analogue to cuprates, an emergent electronic order from vortex halos has been proposed to explain these exotic vortex states, but further microscopic investigations on the high-field vortex state of (TBA$^+$)$_x$FeSe are needed.**

The emergence of superconductivity (SC) fundamentally requires two prerequisites: electronic pairing and macroscopic phase coherence[1,2]. In two-dimensional (2D) SC systems, the occurrence of phase coherence naturally separates from the electronic pairing because strong fluctuations induce an algebraic decay of the phase correlation function with distance[3]. Below the mean-field pairing temperature, the superfluid density is established via the well-established Berezinskii–Kosterlitz–Thouless (BKT) transition[4,5]. SC coherence arises from the binding of vortex–antivortex pairs and is destroyed through the proliferation of free vortices[6].

In quasi-2D layered superconductors, the vortex phase diagram under magnetic fields is strongly affected by the abovementioned BKT physics and exhibits a rich variety of vortex states. A schematic vortex phase diagram for weakly disordered 2D superconductors is presented in Fig. 1a. When a magnetic field ($H$) beyond $H_{c1}$ is applied, the magnetic flux penetrates the superconductor and forms a vortex lattice inside. With increasing $H$, the in-plane electromagnetic interaction between vortices becomes dominant over interlayer Josephson coupling, which drives a crossover of vortices from conventional 3D flux lines to 2D pancake vortices[7–10]. At low fields, a Bragg glass (BG) phase with quasi-long-range order can be stabilized by weak disorder[11,12]. At high magnetic fields, as the role of quantum fluctuations and quenched disorder becomes important, the BG phase melts into a quantum vortex liquid (QVL) phase at finite temperatures[13–15], which is often described as a vortex "slush" phase characterized by a nonohmic current–voltage dependence and a finite resistance $R$ with the applied current $I \to 0$[16,17]. At the zero-temperature limit, this QVL phase is predicted to condense into a dissipation-free 2D vortex glass (VG) in 2D SC systems[9,18]. Alternatively, if strong quantum fluctuations occur within an emergent granular system, the high-field ground state could evolve into an anomalous metallic (AM) state or "failed superconductivity"[19]. Near the upper critical field $H_{c2}$, the interplay between strong Ohmic dissipation and superconducting rare regions may also give rise to a quantum critical region governed by the Quantum Griffiths Singularity (QGS)[20]. As the effects of quenched disorder and quantum fluctuations are thermally suppressed upon heating, an ohmic vortex liquid (VL) would eventually appear at higher temperatures[16,17].

In cuprate superconductors, it is widely believed that the phase fluctuations of Cooper pairs play a crucial role in determining the value of superconducting temperature $T_c$[1,21,22]. In particular, strong electronic correlations foster competition and intertwinements among diverse electronic orders and superconductivity[23–26]. This interplay not only modulates the Cooper pairing strength and disrupts superconducting phase coherence across the doping-dependent superconducting phase diagram, but also critically affect the vortex phase diagram under magnetic fields[13–15,27–30]. Notably, the interplay of competing electronic orders and superconductivity strongly modulates the irreversibility fields[31,32] and drives the emergence of exotic vortex states under high magnetic fields[14,27,28,30], which is beyond the above-discussed paradigms for vortex matter. In charge-ordered cuprates, nondissipative resilient quantum vortex matter with quantum oscillations persisting up to ultrahigh magnetic fields has been observed at extremely low temperatures[28]. Fragile superconductivity, in which both the $T_c$ and critical current density $J_c$ exhibit pronounced suppression relative to their zero-field values, has been proposed to depict this intriguing high-field quantum vortex state[27–29,33]. In this picture, the competing charge order spatially fragments the superconductivity and severely attenuates the macroscopic superfluid stiffness. At present, studies on fragile superconductivity in high magnetic fields are still limited, and further exploration of more high-$T_c$ superconductors is urgently needed.

In this work, we explore the possible fragile superconductivity at high magnetic fields in a quasi-2D FeSe-based superconductor (TBA$^+$)$_x$FeSe. Owing to the additional electron doping into the FeSe layers by intercalating TBA$^+$ cations (tetrabutylammonium, [N(C$_4$H$_9$)$_4$]$^+$) into high-quality FeSe single crystals, the zero-resistivity transition temperature is largely increased from 9 K in pristine FeSe to above 40 K in (TBA$^+$)$_x$FeSe[34,35]. A fully gapped SC state is identified in (TBA$^+$)$_x$FeSe by previous scanning tunneling spectroscopy (STS)[34] and muon spin rotation (μSR) measurements[36]. These results indicate that (TBA$^+$)$_x$FeSe belongs to the family of heavily electron-doped FeSe superconductors, in which the representative superconducting materials include (Li,Fe)OHFeSe[37] and monolayer FeSe/STO thin films[38]. Notably, by the intercalation of TBA$^+$ cations, the interlayer spacing between FeSe layers in (TBA$^+$)$_x$FeSe is expanded to ~ 1.6 nm (see Fig. 1b), which results in a quasi-2D electronic structure with the largest resistivity anisotropy of $\frac{\rho_c}{\rho_{ab}} \sim 10^5$ among all Fe-based superconductors[34,35]. Previous nuclear magnetic resonance (NMR) measurements have revealed persistent superconducting fluctuations well above the zero-resistivity transition temperature[34], suggesting strong phase fluctuations in (TBA$^+$)$_x$FeSe. Here, by performing comprehensive high-field transport measurements up to 33 T, evidence for fragile superconductivity at high magnetic fields is observed in (TBA$^+$)$_x$FeSe.

**Extreme Type II superconductors with large thermal fluctuations**

Fig. 1c presents a typical zero-field $R(T)$ curve for (TBA$^+$)$_x$FeSe. Below 100 K, the in-plane resistance decreases linearly upon cooling. Below 60 K, the $R(T)$ deviates from the $T$-linear behavior, and the slope gradually increases, which is consistent with the pseudogap-like behavior observed in previous NMR experiments and suggests the development of superconducting fluctuations at approximately 60 K[34]. In this quasi-2D superconductor, a BKT-like transition occurs at 46 K (the inset of Fig. 1c), indicating the establishment of superfluid stiffness[6]. As the temperature further decreases, the resistance completely vanishes below approximately 40 K.

Next, we focus on the study of superconducting transitions under different magnetic fields. Isothermal magnetoresistance (MR) curves were traced up to $\mu_0 H_{//c} = 33\ T$ at different temperatures (see Extended Fig. 1a). Through interpolation of the MR data, we reconstructed the $R(T)$ curves under different magnetic fields in Fig. 1d. The irreversibility field $\mu_0 H_{irr}(T)$, corresponding to the melting of vortices, is determined at 0.01% $R_{60K}$ (Fig. 1e). With increasing magnetic field, a 3D to 2D crossover of the vortices occurs at low magnetic fields[8,39–41]. Below a crossover field $\mu_0 H_{cr} = \frac{\phi_0}{d^2 \gamma^2} \sim 0.2\ T$ (where $\phi_0$ is the magnetic flux quantum, $d = 16$ Å is the interlayer spacing, and $\gamma = H_{c2}^{ab}/H_{c2}^{c}$ is the anisotropic ratio), the vortex solid (VS, see Methods) goes through a 3D thermal melting[8]. $\mu_0 H_{irr}(T)$ follows a power–law relationship with temperature: $\mu_0 H_{irr}(T) = \mu_0 H_1 (1 - (\frac{T}{T_1})^2)$. However, deviating from the 3D melting behavior, almost all the irreversibility fields $\mu_0 H_{irr}(T)$ above 0.2 T follow an empirical exponential law $\mu_0 H_{irr}(T) = \mu_0 H_0 \exp\left(-\frac{T}{T_0}\right)$ (Fig. 1e). Such an exponential dependence is typical for quasi-2D superconducting systems dominated by weak bulk pinning and strong surface barriers[42] (e.g., Bi-2212[43] and Hg-1201[44]). This suggests that pancake vortices are pervasive in the $H$-$T$ phase diagram. A considerable anisotropic ratio $\gamma > 60$ is estimated for $(TBA^+)_x FeSe$. As shown in Extended Fig. 1b, by fitting the upper critical fields $\mu_0 H_{c2}^c(T)$ with the 2D Ginzburg–Landau (GL) equation[45]: $\mu_0 H_{c2}^c(T) = \frac{\Phi_0}{2\pi \xi_{ab}^2}\left(1 - \frac{T}{T_c}\right)$, a zero-temperature upper critical field $\mu_0 H_{c2}^c(0) \approx 75\ T$ is estimated. This yields an in-plane superconducting coherence length of $\xi^{ab}(0) \sim 2.1\ nm$. Both $\mu_0 H_{c2}^c(0)$ and $\xi^{ab}(0)$ are comparable with those of (Li,Fe)OHFeSe determined by transport[46] and STS experiments[47].

In the low-$H$ regime, pronounced thermal fluctuations drastically suppress the zero-resistance temperature and lead to a significant broadening of the superconducting transition. The strength of the thermal fluctuations can be quantified by a dimensionless Ginzburg number[48] $Gi = \frac{1}{8}\left(\frac{16\pi^2 \mu_0 \lambda^{ab}(0)^2 \gamma k_B T_c}{\phi_0^2 \xi^{ab}(0)}\right)^2$. Conventional isotropic 3D superconductors, e.g., Nb and Sn, are characterized by minimal Gi values ($\sim 10^{-9}$). Iron-based superconductors, e.g., $RbEuFe_4As_4$, pristine FeSe and $(Ba, K)Fe_2As_2$, etc., generally have Gi values between 0.01 and 0.1[49,50] (see Extended Table 1). However, by using the reported penetration depth $\lambda^{ab}(0) \sim 240 - 500\ nm$[36,51], the Gi value of

(TBA$^+$)$_x$FeSe is estimated to be extremely large (> 320). This large value of Gi resides within the characteristic range of Bi-2212 (Gi = 329)[48] and Hg-2223 (Gi = 112)[52] (Fig. 1d), indicating massive thermal fluctuations in (TBA$^+$)$_x$FeSe. In addition, the Ginzburg–Landau parameter $\kappa^c = \frac{\lambda^{ab}}{\xi^{ab}}$ is estimated to be 110 - 240. All these results suggest that (TBA$^+$)$_x$FeSe is an extreme type II superconductor.

**Exotic vortex states at high magnetic fields**

At low temperatures and high magnetic fields, the $R$(T) curves exhibit 2D VG characteristics as shown in Fig. 2a. Below 2.2 K, the $R$(T) data can be well fitted by a power law relation[17,53,54]: $R(H,T) = R_0(H)T^{\alpha(H)}$. Fig. 2b displays the field dependence of the fitting parameters $R_0(H)$ and $\alpha(H)$. Notably, above $\mu_0 H_{VG} \sim 31$ T, $R_0(H)$ rapidly increases with increasing magnetic field, and the exponent $\alpha(H)$ slightly decreases, which indicates a 2D VG ground state at the zero-temperature limit. A similar 2D VG state has been also observed in cuprate superconductors[17,53]. In that case, when $\alpha(H)$ approaches zero at high magnetic fields, the 2D VG at high magnetic fields eventually evolves into a metallic or insulating state via quantum phase transitions (QPTs)[17,53,54], which is beyond the focus of superconducting ground state in the present study.

At high magnetic fields, the 2D VG with resistance $R = \lim_{I \to 0} \frac{V}{I} = 0$ melts above 0 K, where quantum fluctuations overcome the pinning forces from quenched disorder and create a nonohmic glassy vortex liquid. To explore the effect of quantum fluctuations on the high-field vortex states, we applied excitation currents spanning three orders of magnitude, from $2\,\mu A$ to $1000\,\mu A$ ($J = 5 \times 10^{-5} - 2.5 \times 10^{-2}\, A/mm^2$), to measure the isothermal magnetoresistances up to 33 T.

As shown in Fig. 3, the resistance measured at higher excitation currents ($I = 200\, or\, 1000\, \mu A$) is significantly greater than that measured at lower currents ($I = 2\, \mu A\, or\, 20\, \mu A$) at low temperatures. The increase in resistance is mostly remarkable at the lowest temperature. With increasing temperature, this nonohmic current effect gradually diminishes and vanishes at 15 K, where the isothermal MR lines measured at $20\, \mu A$ and $1000\, \mu A$ nearly overlap with each other. We identify a temperature-dependent characteristic field $H_{AVL}(T)$ defined by the point where these MR curves converge (see insets in Fig. 3 d~f). This $H_{AVL}(T)$ demarcates an anomalous vortex–liquid regime characterized by pronounced nonohmic transport behavior, which appears to extend up to ultrahigh magnetic fields.

In addition to current effects on nonohmic resistive states, we find that the irreversibility fields $\mu_0 H_{irr}(T)$ can also be significantly suppressed to lower magnetic fields by modest excitation currents. This current-induced suppression of $\mu_0 H_{irr}(T)$ also becomes most prominent at the lowest temperature, where the discrepancy of the irreversibility fields $(\Delta(\mu_0 H_{irr}(T)) = \mu_0 H_{irr}^{20\mu A}(T) - \mu_0 H_{irr}^{1000\mu A}(T))$ reaches 5 T at $T = 1.75\,K$ as shown in Fig. 3a. At low temperatures, a critical-current-suppressed superconducting state is induced by magnetic fields (see Extended Fig. 2a). Notably, the Joule heating effects from the excitation currents can be experimentally excluded in our case (see details in Methods). This current effect on superconducting state strongly suggests a fragile superconductivity at high magnetic fields. More discussions on this issue will appear later.

**Phase-fluctuating vortex state at finite temperatures**

Furthermore, we study the temperature-dependent evolution of vortex states by measuring the Hall resistance (HR, $R_{xy}$) and longitudinal magnetoresistance (MR, $R_{xx}$) simultaneously. Figs. 4a - d display $R_{xy}$ and $R_{xx}$ traces acquired with an excitation current of $I = 1000\,\mu A$ at 1.75 K, 5 K, 12 K, and 25 K, respectively. A key observation is that the Hall responses remain vanishingly small well into the magnetic field at which the longitudinal resistances have already become significant. The HR lines collected at 1.75~35 K are plotted in Fig. 4e, and the onset magnetic fields of the zero-Hall state are also marked. As shown in Fig. 5b, a particle–hole symmetry preserved phase-fluctuating vortex state with $R_{xx} > 0$ and $R_{xy} \sim 0$ is demarcated on the phase diagram.

The physical origin of this zero-Hall resistive state remains a subject of active debate. As we discussed before, AM states with $R_{xx} > 0$ and $R_{xy} \sim 0$ have been widely found in quasi-2D granular superconductors[19]. The AM state is characteristically associated with suppressed Hall responses[55,56], saturation of resistivity at $T \to 0\,K$[19,20], power-law magnetoresistance scaling[57] and the absence of cyclotron resonances[58]. A Bose metal scenario with particle–hole symmetry was proposed years ago to explain the AM state, wherein Cooper pairs mediate dissipations via glassy superconducting islands connected by random Josephson couplings[59–61]. An alternative framework involves a Bosonic topological insulator[62] driven by quantum fluctuations. Furthermore, additional theories, such as quantum tunneling of vortices at low temperatures[63], composite fermion models[64] and stripe order[65,66], have been developed. In contrast to Bose metal scenario, the phase-fluctuating vortex state discussed here is not a real ground state. The ground state is still a fragile superconducting state in the present study. In the following section, we will discuss the specific mechanism for the zero-Hall resistive state in $(TBA^+)_x$FeSe.

**Discussions**

We summarize all the transport characteristics in the phase diagram presented in Fig. 5b. In the high-$H$ regime, the irreversibility fields obtained by 20 and 1000 $\mu A$ delineate a "resilient" superconducting state that is highly sensitive to external currents. While one might attribute this low-critical-current superconductivity to the reduced rigidity of quasi-2D pancake vortices or the effects of weak vortex pinning, such mechanisms typically yield a relatively uniform response across the magnetic field and temperature range. In stark contrast, the fragile superconducting state observed here is explicitly field-induced, becoming increasingly prominent in the high-$H$ and low-$T$ regimes. This suggests a mechanism beyond the simple disordered pinning effect and the dimensionality of the vortex. The markedly reduced superconducting $T_c$ and critical current $j_c$ bear a striking resemblance to the fragile superconductivity previously proposed in charge-ordered cuprates[33]. Quantum fluctuations, which become increasingly relevant in the low-$T$, high-$H$ limit, might be instrumental in the emergence of this fragile superconducting state. Notably, extensive studies have suggested that such fragile SCs, along with nonohmic vortex liquids, emerge exclusively in cuprates hosting long-range charge orders[27].

As theorists posited, in a globally CDW-dominated system, subdominant superconductivity emerges locally on CDW dislocations[33]. The critical current of the entire system is limited by the Josephson current, which can be sustained by the weakest links between the nucleated "superconducting halos"[33]. From an alternative perspective of vortex physics, with increasing magnetic fields, when the charge-order coherence length $\xi_{CO}$, which thrives in the vortex cores[24,67–69], exceeds the intervortex spacing, then the emergent charge order may interfere with vortex dynamics, further degrading the phase coherence and current-carrying capacity[27].

Recently, accumulating evidence has indicated the existence of charge orders and pair density waves (PDWs) in heavily electron-doped FeSe-derived superconductors, including a checkboard charge order around Fe defects in (Li, Fe)OHFeSe films[70] and an incommensurate PDW surviving at the domain walls of monolayer Fe(Se, Te)/STO films[71]. Motivated by these results, we speculate that the observed exotic transport phenomena are intimately linked to these competing or intertwined electronic orders.

We propose the following scenario at low temperatures. An incipient electronic order emerges within the vortex halos. As the vortex density increases, this proliferated electronic order disrupts global superconductivity. Specifically, near vortex cores, once the superconductivity couples to this electronic order, the phase of the order parameter varies randomly from one vortex halo to another. The global superconductivity is fragmented to phase-randomized superconducting "puddles" interconnected via weak Josephson couplings[33] (see Fig. 5a). At high magnetic fields, phase coherence persists locally within individual puddles, whereas the global coherence becomes fragile[33]. The reduced $j_c$ is mediated by these weak Josephson couplings[33] between the puddles. Thus, external excitations or thermal/quantum fluctuations can easily eliminate the global coherence and transform the fragile SC into the phase-fluctuating vortex state (Extended Fig. 2a).

The origination of the phase-fluctuating vortex state warrants further discussion. Phenomenologically, this undetectably small Hall response implies that the density of vortices comoving with the superfluid is negligible by strong pinning[72]. Then, is it plausible to ascribe the origination of this dissipation to anomalous vortex motion? A recent study on FeSe/SrTiO$_3$ (STO) thin films attributed their anomalous metallic behaviors to the quantum tunneling of vortices[63]. Additionally, the motions of vortex dislocations in the 2D hexatic vortex phase have been reported to yield negligible Hall responses[73]. Nevertheless, these two mechanisms do not fully capture our observations.

Typically, the AM states observed in 2D superconducting films are characterized by a finite, saturating resistance as $T \rightarrow 0$, representing a stable Ohmic ground state at magnetic fields exceeding the zero-temperature irreversibility field, $\mu_0 H_{irr}(0\ K)$. In contrast, the zero-Hall state in (TBA$^+$)$_x$FeSe manifests as a nonohmic transport regime inconsistent with a stable metallic ground state. This state appears to collapse above $\mu_0 H_{irr}(0)$ as the temperature approaches 0 K, precisely where the quantum tunneling of vortices is expected to dominate. The phase-fluctuating vortex state evolves to an anomalous vortex liquid at higher $H$, distinguished by the reemergence of a finite Hall response (see Extended Fig. 2a). This implies that the phase-fluctuating vortex state here is not an AM ground state and may not be governed by the quantum tunneling of vortices. In addition, a resistive plateau, hallmark of the Hexatic phase[73], is absent on the $R$(T) or MR curves of (TBA$^+$)$_x$FeSe, which helps to preclude the vortex Hexatic phase scenario. Remarkably, the onset fields of the phase-fluctuating vortex state in (TBA$^+$)$_x$FeSe exhibit an independence of the excitation currents over a broad range ($20 - 1000\ \mu A$) (Extended Fig. 3). The relative robustness of this zero-Hall, nonohmic region is indicative of strong global vortex pinning. This raises the intriguing question of how such strong pinning emerges in a quasi-2D pancake vortex system, where vortex pinning is conventionally expected to be weak.

We believe that this phase-fluctuating state is an extended fragile superconductivity persisting at elevated temperatures with an exceptionally low critical current. It outlines a nearly frozen vortex liquid region on the $H$-$T$ phase diagram that precedes the fully unpinned regime. Within this framework, we propose that emergent granularity is the key driver of the enhanced pinning, and the dissipation plausibly results from vortex-like phase fluctuations across the weak links within an intrinsically granular superconducting state under $H_{//c}$[19, 54].

We suggest that the strong pinning and granular superconductivity is not driven primarily by extrinsic factors such as strong crystalline defects or nonuniform distributions of chemical concentrations, given the sharp SC transitions in the AC and DC magnetizations and narrow NMR spectra in the normal state (Extended Fig. 4). Compared with the (Li, Fe)OHFeSe and Fe(Se, Te)/STO films, our (TBA$^+$)$_x$FeSe possesses more structurally intact and cleaner FeSe layers[74]. Furthermore, this vanishing-Hall resistive state exhibits robust reproducibility across multiple (TBA$^+$)$_x$FeSe samples with different residual resistance ratios (RRR $= 5 - 20$) (Extended Fig. 5), which helps to rule out impurities as its origin as well. Consequently, we propose that this emergent granular superconductivity is rooted in the intrinsic interplay of competing orders, e.g., CDW and SC, under magnetic fields[19]. Our data indicate that this inhomogeneous superconducting state already emerges at relatively low magnetic fields and potentially persists deep into the high-field regime. In the extremely high-$H$ and low-$T$ regimes, strong quantum fluctuations or a globally established electronic order might suppress both the superconducting phase coherence and the particle–hole symmetry.

In summary, our high-field transport measurements reveal an intricate vortex phase diagram in the quasi-2D FeSe-based superconductor (TBA$^+$)$_x$FeSe. The fragile superconductivity and phase-fluctuating vortex state observed here closely parallel those of cuprate superconductors, suggesting universal high-field vortex physics in high-$T_c$ superconductors. This universal high-field vortex physics might provide a new window for understanding the establishment of phase coherence in high-$T_c$ superconductivity. In addition, to unambiguously identify the emergent electronic order from the vortex halo, further microscopic spectroscopic investigations, e.g., NMR, STS, resonant inelastic X-ray scattering (RIXS) and high-field quantum oscillation measurements, are urgently needed.

**Main Figures**

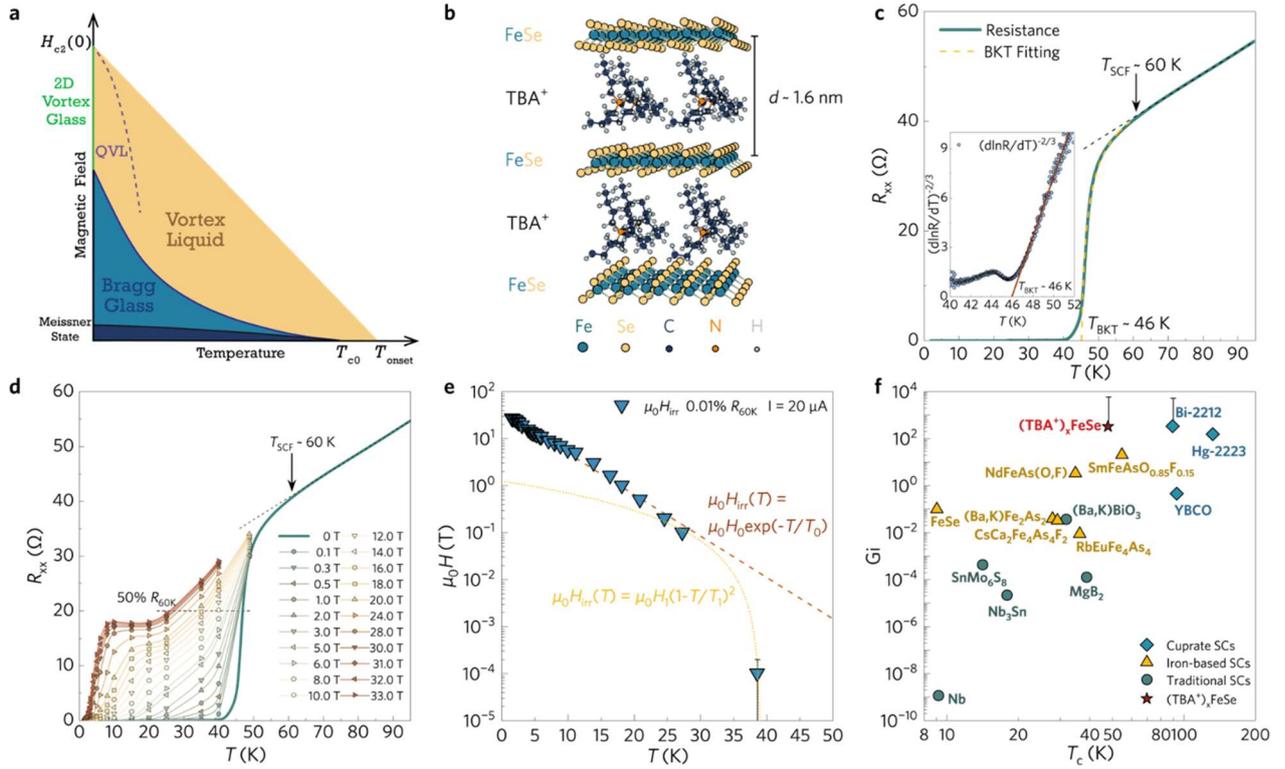

**Fig. 1| Superconducting transition under magnetic fields in $(TBA^+)_xFeSe$. a,** A schematic vortex phase diagram of weakly disordered quasi-2D superconducting systems with pronounced quantum/thermal fluctuations and superconducting fluctuations under magnetic fields perpendicular to superconducting planes. **b,** A schematic crystal structure of $(TBA^+)_xFeSe$. The distance between the superconducting FeSe layers is estimated as $d \sim 1.6\ nm$ via X-ray diffraction[34,75]. **c,** Zero-magnetic-field $R(T)$ curve. Superconducting fluctuations (SCFs) start to emerge below 60 K. The dashed line is the fit of the Halperin–Nelson equation $R(T) = R_0 \exp\left[-b/(T-T_{BKT})^{1/2}\right]$. A BKT transition temperature $T_{BKT} = 46\ K$ is obtained, below which the superfluid stiffness is established. Insert: The temperature-dependent differential resistance of sample #1 in the form of $(d\ln R/dT)^{-2/3}$. **d,** The temperature-dependent resistance $R(T)$ at various magnetic fields up to 33 T. The data points are interpolated from magnetoresistance curves down to 1.55 K (see Extended Fig. 1). Below 20 K, we applied $I = 20\ \mu A$ for the measurements. The other high-temperature data were traced with $I = 1000\ \mu A$. **e,** Irreversibility points $\mu_0 H_{irr}(T)$ (denoted at 0.01% $R_{60K}$) captured by $I = 20\ \mu A$. The points here are composed of data on Samples #1 and #4 (see Extended Fig. 6). The $\mu_0 H_{irr}(T)$ points below 0.2 T follow a power-law relationship with temperature: $\mu_0 H_{irr}(T) = \mu_0 H_1 (1-(T/T_1)^2)$ with $H_1 \sim 1.25\ T$ and $T_1 \sim 39\ K$. Above 0.2 T, it can be well fitted by $\mu_0 H_{irr}(T) = \mu_0 H_0 \exp(-T/T_0)$ with $H_0 \sim 36\ T$ and $T_0 \sim 4.9\ K$. The error bars represent the uncertainties in determining the irreversibility fields. **f,** Ginzburg–Levanyuk numbers of several series of superconductors. The star symbol represents the low-limit estimation of the Gi number of $(TBA^+)_xFeSe$, and the error bars denote the upper limit of the estimations. The explicit superconducting parameters are summarized in the Extended Table 1.

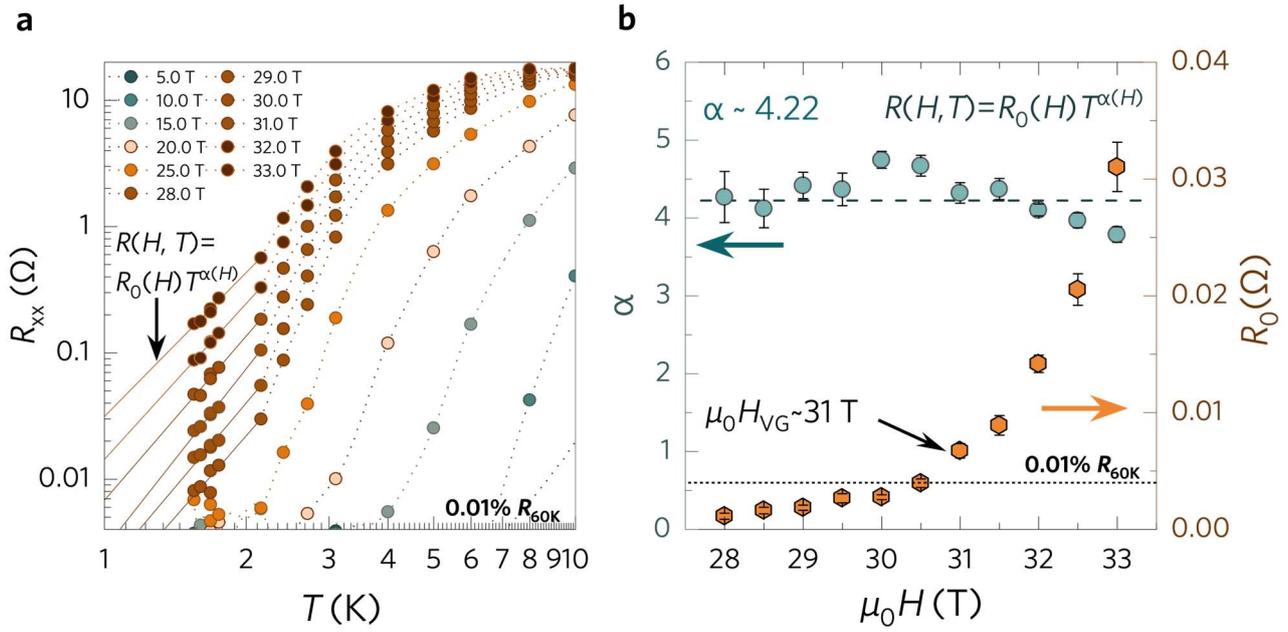

**Fig. 2| Evidence for a 2D vortex glass state at the zero-temperature limit. a,** A partially detailed view of **Fig. 1d**, focusing on the low-$T$ and high-$H$ regimes. The solid lines represent power-law fittings $R(H,T) = R_0(H)T^{\alpha(H)}$ for data points between 1.55 K and 2.2 K at high magnetic fields. **b,** Magnetic field dependence of the fitting parameters $R_0(H)$ and $\alpha(H)$ with fitting error bars. The $R_0(H)$ starts to be prominent above a characteristic magnetic field $\mu_0 H_{VG} \sim 31\ T$, corresponding to the formation of 2D vortex glass at 0 K.

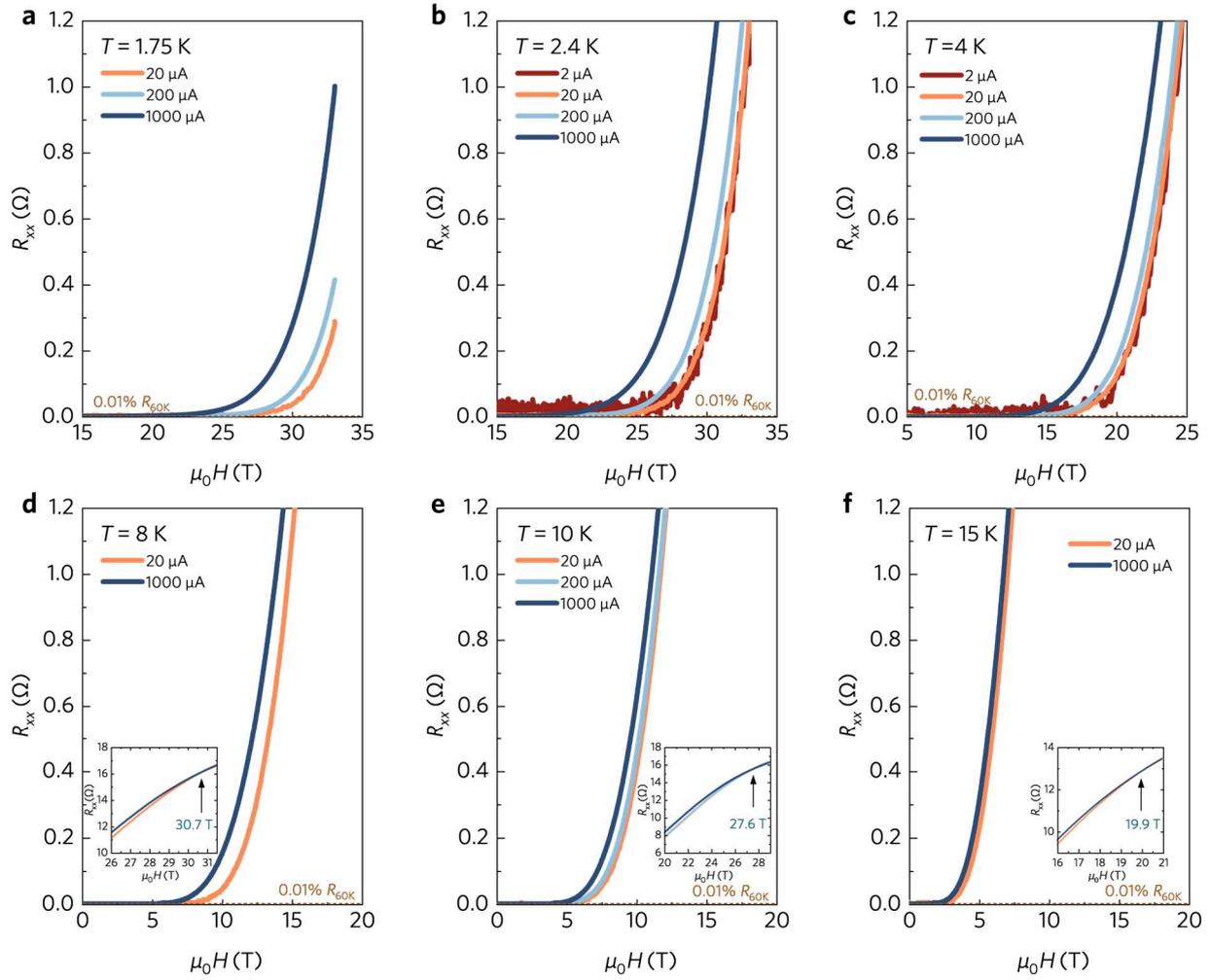

**Fig. 3| Field-dependent superconducting transition under different currents. a~f**, With the application of excitation currents spanning three orders of magnitude from 2 $\mu A$ to 1000 $\mu A$, the magnetoresistances were traced at 1.75 K, 2.4 K, 4 K, 8 K, 10 K and 15 K, respectively. Modest currents can significantly affect both the coherent and the resistive state. The inserts in **d~f** show that the resistive lines merge at some specific magnetic fields, indicating the elimination of current effects.

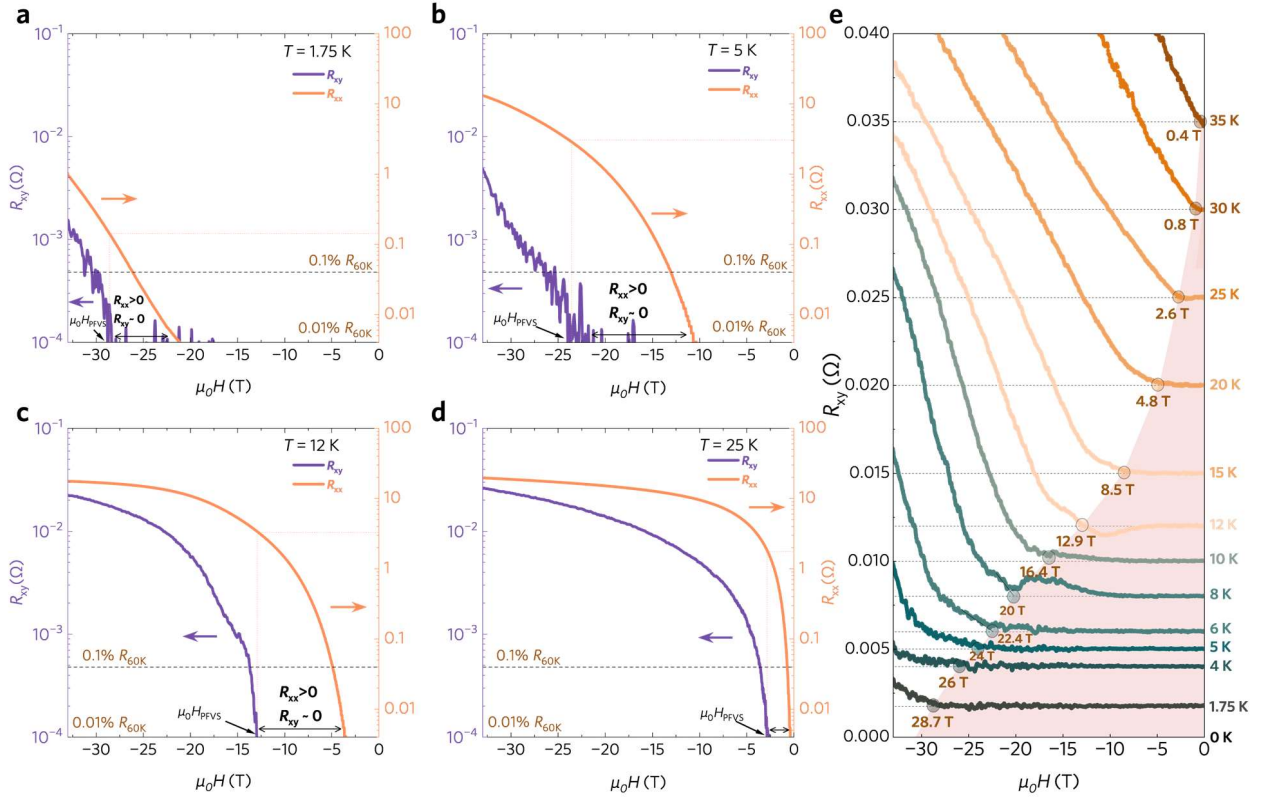

**Fig. 4| Evidence for a phase-fluctuating vortex state (PFVS) at different temperatures. a~d**, Isothermal magnetic field-dependent Hall resistances $R_{xy}(H, T)$ and longitudinal resistances $R_{xx}(H, T)$ at -33 T ~ 0 T were measured at 1.75 K, 5 K, 12 K and 25 K, respectively. Considering the negative Hall coefficient $R_H$, we plot negative magnetic fields here for clarity of data presentation. For all the Hall and MR measurements, the applied excitation I = 1000 μA. The 0.1% and 0.01% $R_{60K}$ lines are guides for visual inspection. We denote the onset of the phase-fluctuating vortex state as the Hall resistance decreases below the resolution capacity of our instruments (~$10^{-4}$ Ω). **e**, A collection of Hall resistance $R_{xy}(H, T)$ and the onset magnetic fields of the phase-fluctuating vortex state below 35 K. Each of the Hall resistance lines is offset according to its temperature. The pink patch outlines the boundary of the PFVS.

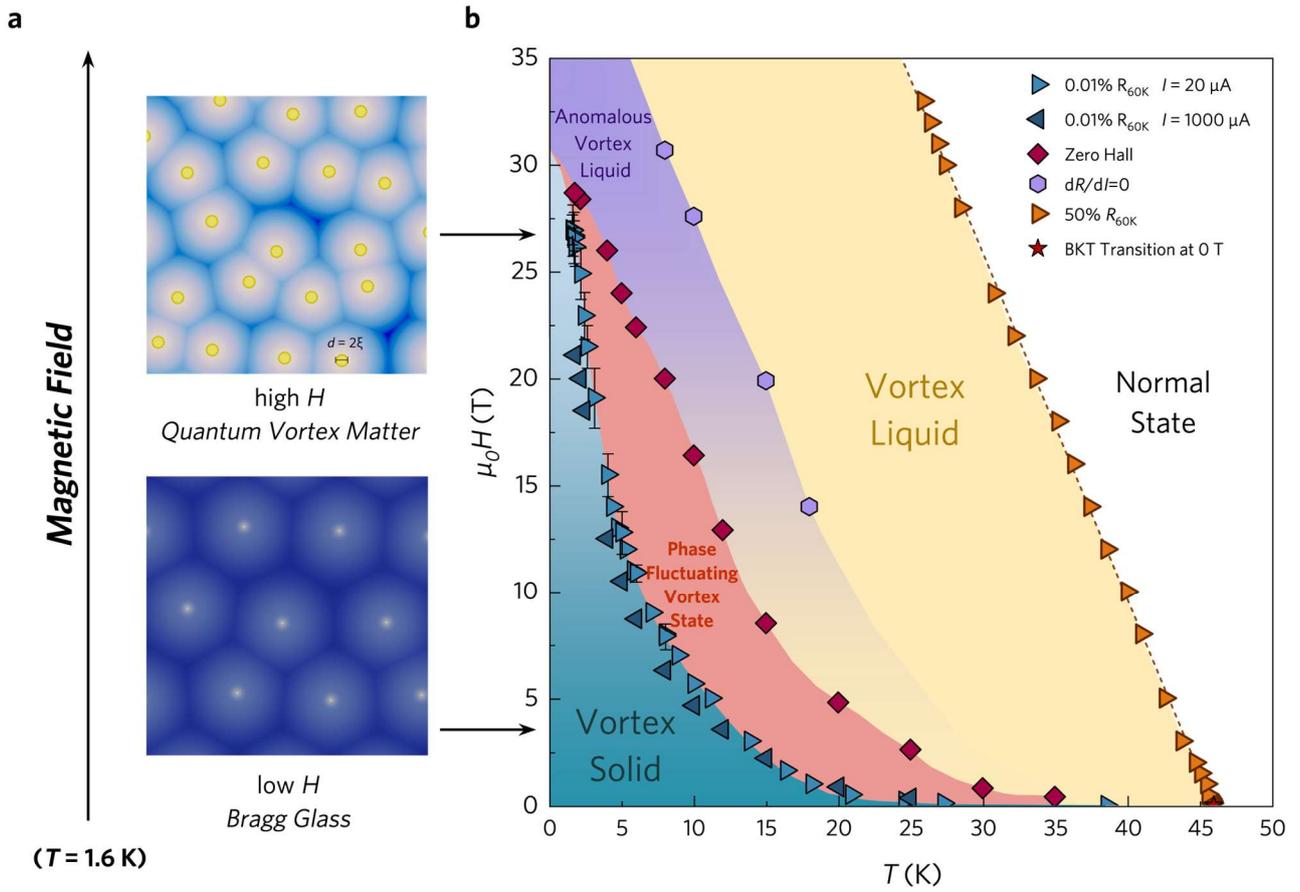

**Fig. 5| *H–T* phase diagram of the vortex state with emergent electronic order from the vortex halo. a,** Schematic diagram of vortices with emergent electronic order. The bright yellow spots represent vortex cores, whereas the dark blue regions denote the superconducting background. White patches around cores indicate the coexistence areas of superconductivity and the competing electronic state. At low magnetic fields, vortices are widely spaced with weak mutual interactions, leading to quasi-long-range hexagonal symmetric Bragg glass. In this regime, the interplay between the competing electronic states and superconductivity remains weak. As the magnetic field increases, the vortex density increases, enhancing the interactions among vortices. Under the influence of disorder, the vortex lattice loses both its sixfold symmetry and long-range translational order, adopting a vortex-glass-like configuration. Simultaneously, the emergent electronic state from the vortex cores becomes more pronounced at high fields. This competing electronic order exists with superconductivity and modulates it around the vortices. Overall, the superconducting state is segmented into granular regions and forms a quantum vortex matter at high magnetic fields. **b,** *H-T* phase diagram of (TBA$^+$)$_x$FeSe determined by transport measurements. The vortex solid represents a nondissipation transport regime with excitation currents $I \rightarrow 0$. The phase-fluctuating vortex state (PFVS) denotes a state with marked resistances but negligible Hall responses. The anomalous vortex liquid is a finite resistive state ($R > 0$ as $I \rightarrow 0$), where the excitation currents can significantly affect the resistance. The vortex liquid is the ohmic region where currents ($I < 1000\ \mu A$) can hardly influence the resistance, whereas vortices still exist for electronic pairing. The upper critical field $\mu_0 H_{c2}^c(T)$ is extracted from the $R(H_{c2}, T) = 50\% R_{60K}$ points, delineate an underestimated superconducting fluctuating regime. The irreversibility points collected at 20 $\mu A$ and 1000 $\mu A$ are plotted to demonstrate a "resilient" fragile superconducting state in the high-*H* and low-*T* regimes.

## Methods

**Synthesis of the (TBA$^+$)$_x$FeSe samples:** High-quality pristine FeSe single crystals were synthesized via the chemical vapor transport method[76] and exhibited a superconducting transition temperature ($T_c$) of approximately 9 K. A single-crystal of mass 0.94 mg was selected for chemical intercalations. The electrolyte was prepared by dissolving tetrabutylammonium bromide (TBAB, Aladdin, 99.0%) (3.00 g) in *N,N*-dimethylformamide (DMF, Innochem, 99.9%, extra dry, H$_2$O < 50 ppm) (10 mL). The pristine FeSe crystal, attached to an indium wire, was used as the cathode, whereas a silver strip (99.99% purity) served as the anode. The electrodes were immersed in the prepared electrolyte, and an electrochemical current of 20 μA was applied for 10 to 15 hours until the electrochemical potential was reduced from -0.3 V to -2 V. This process resulted in the successful preparation of the intercalated superconductors (TBA$^+$)$_x$FeSe. An electron probe X-ray microanalyzer (EPMA) was used to estimate a stoichiometry of x ~ 0.2 ions per FeSe unit cell, corresponding to (TBA$^+$)$_{0.2}$FeSe[34].

**Transport measurements at high magnetic fields:** We cleaved (TBA$^+$)$_x$FeSe into a slice with a size of approximately 1 mm×0.4 mm×0.1 mm. Because the sample is extremely air sensitive, we sealed the sample along with the electrodes into a Quantum Design HPC-33 piston-type pressure cell, using Daphne 7373 oil as the pressure transmitting medium to isolate the water and oxygen. The longitudinal and Hall voltages were collected simultaneously by a six-terminal electrode via two lock-in amplifiers (SR830) with the same AC currents (2–1000 μA, ~13 Hz) generated by a Keithley 6221 current source. The high-magnetic-field transport measurements were performed on a water-cooled magnet (WM5) at the Steady High Magnetic Field Facilities (SHMFF), High Magnetic Field Laboratory, CAS. The Hall resistance ($R_{xy}$) is obtained via a standard anti-symmetrization procedure $R_{xy}(\mu_0 H) = \frac{V(+\mu_0 H) - V(-\mu_0 H)}{2I}$. After intercalation, the FeSe layers are covered with TBA$^+$ organic ions, which makes the macroscopic structure of the layered sample uncompacted and easy to exfoliate. Notably, the resistivity is hard to be precisely determined because of the uncertainty of the effective conductive thickness. As a result, the absolute values of the Hall coefficient and resistivity are not shown in this paper. According to the geometric structure of the six-terminal electrodes and the sample, its resistivity at 60 K $\rho_{60K}$ is estimated at most as 0.4 Ω·cm, which is believed to couple some dissipations from organic ions layers along the c axis. The current density $J$ is estimated as $5 \times 10^{-4}\ A/mm^2$ for the applied current $I = 20\ \mu A$ and $2.5 \times 10^{-2}\ A/mm^2$ for the $I = 1000\ \mu A$ excitation. Sample #1 has a residual resistance ratio (RRR) of ~ 5, which is not directly shown in the text. Sample #2 and Sample #3 have RRRs of ~ 5 and 20, respectively.

**The determination of critical magnetic fields:** The irreversibility line in Fig. 1d and 5b is composed of data points collected from sample #1 at a high-field water-cooled 33 T magnet and points from sample #4, which has a size close to that of sample #1, at a 14 T DynaCool (Quantum Design). The applied currents are both 20 $\mu A$. The irreversibility line criteria for both data series are the same: 0.01% resistance at 60 K. The R-T curves at DynaCool are plotted in Extended Fig. 6. The vortex solid phase is defined by the vanishing resistance as $I \to 0$[17]. Experimentally, its boundary is delineated via irreversibility points collected with 20 $\mu A$ excitations, a current sufficiently small to approach the $I \to 0$ limit yet large enough for the signal strength. We determine the upper critical fields $\mu_0 H_{c2}(T)$ as the temperature-dependent magnetic fields at which the resistance decreases to 50% of $R_{60K}$ at zero field. However, as we mentioned above, the real superconducting fluctuations emerge just below 60 K, which is beyond the upper-critical-field criterion of 50% of $R_{60K}$. Since the needed magnetic fields for tracking superconducting fluctuations are incredibly large and beyond the magnet capability. Here, we only use the $\mu_0 H_{c2}(T)$ to represent the vortex liquid regime, characterizing an underestimated superconducting fluctuating regime. The "normal state" in Fig. 5b is believed to host some superconducting fluctuations as well.

**Preclusion of the current heating effect:** We compared the isothermal magnetoresistance curves measured at 1.7 K via $I = 1000\ \mu A$ with the resistance at 2.4 K via $I = 20\ \mu A$. The zero resistance states under these two conditions vanish at almost the same magnetic field. If the 1.7 K trace was heated to 2.4 K by a 1000 $\mu A$ current, we would expect an overall overlap of these two magnetoresistance curves[27]. However, we note that they only cross at a single point $\mu_0 H \approx 30\ T$. The cases at 5 K, 1000 $\mu A$ and 6 K, 20 $\mu A$ are also compared (see Extended Fig. 7). These two lines cross at a single point $\mu_0 H \approx 13.5\ T$. Larger currents ($I = 1000\ \mu A$) are likely to produce greater resistance due to the heating power $P = RI^2$, which could accelerate increases in the resistance if the heating power is significant. In contrast, our data show that the MR lines, which share the same irreversibility field, only coincide at one resistive point. The $I = 1000\ \mu A$ lines display lower resistance values than those with $I = 20\ \mu A$ at higher magnetic fields. This suggests that the observed anomalous nonohmic electronic transport of the sample is unlikely to be due to temperature disturbances. Here, we estimate the practical values of Joule heating effects caused by the applied excitation currents and eddy currents induced by changing magnetic fields. At low temperatures below 5 K, the heating power of the excitation current is estimated to reach $10^{-7 \sim 5}\ W$ with $I = 1000\ \mu A$. The heating power by eddy currents on the sample is estimated as $P = I^2 R = V^2/R = (d\Phi_B/dt)^2/R = A^2(dB/dt)^2/R = 1.6 \times 10^{-4 \sim 2}\ pW$ by considering the area of the sample as 0.4 $mm^2$, the changing rate of the magnetic field as 0.1 Tesla/s and the resistance as $R = 0.1 \sim 10\ \Omega$. The heating power of eddy currents on the pressure cell is estimated to reach $5 \times 10^{-7 \sim 5}~W$. However, the cooling power of the VTI is approximately $1~W$ at 2 K, approximately five to seven orders of magnitude greater than the current heating power. The drifts of temperatures in measurements can also be precluded. At any temperature before we started tracking the data and turned on the magnetic field, we kept the temperature steady for at least 15 minutes. In conclusion, the temperatures remained stable during the measurements of magnetoresistance.

**The resistive hump and the quantum critical points at high magnetic fields**

In Fig. 1d, a hump-like feature on the $R(T)$ curve gradually emerges with increasing magnetic field in the intermediate temperature range. A partially detailed view of the hump is shown in the Extended Fig. 8a. Similar field-induced humps have been reported in various 2D superconducting thin films and unconventional layered superconductors. These resistive humps are attributed to a crossover from an Ohmic VL to a glassy VL[77,78] or are interpreted as manifestations of an "exotic vortex liquid"[79]. An alternative interpretation is that the hump originates from the coupling of the $c$-axis dissipation by the motions of nonrigid vortex fluxes during in-plane resistance measurements[80].

However, we propose that the $R(T)$ hump observed in $(TBA^+)_x FeSe$ signifies quantum phase transitions (QPTs). Analogous to observations in La-214 systems[17,53], the hump evolves to peak-like ($dR/dT = 0$) behavior above 31 T, potentially marking a precursor to a superconducting-insulator transition[53] (SIT) or a superconducting-metal transition[17] (SMT) at higher magnetic fields. A quantum critical point (QCP) identified near $\mu_0 H^* \sim 31$ T can be captured by the crossing point of the MR curves at 10 K and 15 K (see Extended Fig. 8). The inset of Extended Fig. 8 shows that additional crossing points of MR curves are expected at lower temperatures, anticipating a multistage QPT or the emergence of a quantum Griffiths singularity above 31 T.

However, the identification of a QCP at 31 T does not explicitly dictate a single phase transition from a Bragg glass to a 2D vortex glass, which is the scenario previously established in $La_{2-x}Sr_xCuO_4$[53]. In $(TBA^+)_x FeSe$, the regime extending above 31 T may instead represent a broad critical region governed by multiple QCPs or arguably by a quantum Griffiths singularity. Consequently, the onset field of the 2D vortex glass behavior $\mu_0 H_{VG} \sim 31$ T might be accidentally coincidence with the crossing point of the magnetoresistances between 10 K and 15 K rather than having a common microscopic origin. This suggests that the correlation between the 2D vortex glass state and the observed QCPs is complex, and further investigations are needed to clarify their intrinsic relationship.

**Data availability**

The data that support the findings of this study are available from the corresponding author upon request.

**Code availability**

The codes that support the findings of this study are available from the corresponding author upon request.

**Acknowledgments**

We are thankful for the valuable discussion with Ziji Xiang and Zhenzhong Shi. This work is supported by the National Key R&D Program of the MOST of China (Grant No. 2022YFA1602601), the National Natural Science Foundation of China (Grants No. 12494592, 12034004, 12161160316, 12325403, 12488201), the Chinese Academy of Sciences under contract No. JZHKYPT-2021-08, the CAS Project for Young Scientists in Basic Research (Grant No.YBR-048), the Innovation Program for Quantum Science and Technology (Grant No. 2021ZD0302800), and the Fundamental Research Funds for the Central Universities (No. WK9990000110). This work was partially carried out at Instruments Center for Physical Science, University of Science and Technology of China.

**Author contributions**

T.W. and X.C. conceived the research project and coordinated the experiments. X.L. grew the high-quality single crystal samples with the help of B.K. and M.S.. X.Y. and J.C. performed the transport measurement under high magnetic field with the help of C.X., J.Z., J.L., K.L., S.L. and D.Z.. X.L. and T.W. analyzed the data and wrote the paper with input from all the authors.

**Extended Tables and Figures**

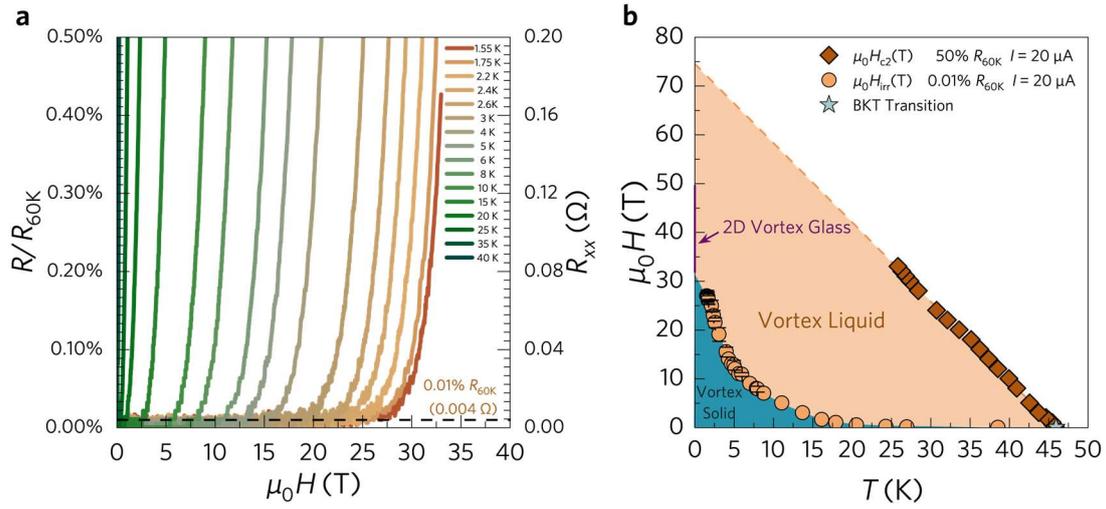

**Extended Fig. 1| The magnetoresistance up to 33 T unveils a broad range of vortex liquid. a.** Isothermal magnetoresistance curves measured under 0 - 33 T magnetic fields at temperatures down to 1.55 K. Below 20 K, we applied $I = 20\ \mu A$ for the measurements. The other high-temperature MR lines were traced with $I = 1000\ \mu A$. The resistive data are normalized by resistance at 60 K and 0 T. The dashed lines are visual guides for the criterion of irreversibility lines (0.01% $R_{60K}$). **b.** Vortex phase diagram of $(TBA^+)_x FeSe$ for $H_{//c}$. We use points $R(\mu_0 H_{c2}, T) = 50\% R_{60K}$ to represent the upper critical field $\mu_0 H_{c2}^c(T)$. The dashed line is the 2D Ginzburg–Landau fit $\mu_0 H_{c2}^c(T) = \frac{\Phi_0}{2\pi \xi_{ab}^2}\left(1 - \frac{T}{T_c}\right)$. The in-plane superconducting coherence length $\xi^{ab}(0) \sim 2.1\ nm$ and an upper critical field $\mu_0 H_{c2}^c(T) \sim 75\ T$ are obtained. A 2D vortex glass phase is predicted above 31 T. A broad dissipative regime with quantum fluctuations is expected to be observed at ultrahigh magnetic fields and low temperatures, similar to Bi-2212[2].

**Extended Table 1| Giant Ginzburg–Levanyuk number in (TBA$^+$)$_x$FeSe compared with several typical superconductors.**

| Superconductors | $\xi_{ab}(0)$ (nm) | $\lambda_{ab}(0)$ (nm) | $\gamma$ | $T_c$ (K) | $Gi$ |
|---|---|---|---|---|---|
| RbEuFe$_4$As$_4$ | 1.46 | 98 | 1.7 | 36.5 | 8.54×10$^{-4}$ |
| FeSe | 3.4 | 357 | 4 | 9.1 | 9.55×10$^{-2}$ |
| SnMo$_6$S$_8$ | 2.8 | 133 | 1 | 14.2 | 4.13×10$^{-4}$ |
| Nb$_3$Sn | 3.4 | 61.7 | 1 | 18 | 2.08×10$^{-5}$ |
| Nb | 28.6 | 21.3 | 1 | 9.25 | 1.10×10$^{-9}$ |
| MgB$_2$ | 10 | 50 | 5 | 39 | 1.22×10$^{-4}$ |
| (Ba,K)BiO$_3$ | 3 | 280 | 1 | 32 | 3.59×10$^{-2}$ |
| (Ba,K)Fe$_2$As$_2$ | 1.6 | 140 | 2.5 | 28 | 3.8×10$^{-2}$ |
| YBa$_2$Cu$_3$O$_7$ | 1.4 | 75 | 7.8 | 93.7 | 0.44 |
| NdFeAs(O,F) | 2.4 | 270 | 7.5 | 35 | 3.26 |
| HgBa$_2$Ca$_2$Cu$_3$O$_{8+\delta}$ | 1.3 | 133 | ~30 | 133 | 151.13 |
| CsCa$_2$Fe$_4$As$_4$F$_2$ | 2.29 | 98.6 | 6.3 | 29.4 | 3.17×10$^{-2}$ |
| SmFeAsO$_{0.85}$F$_{0.15}$ | 0.8 | 189 | 8 | 55 | 19.79 |
| Bi$_2$Sr$_2$CaCu$_2$O$_{8+\delta}$ | 1.1 | 140 | 50-200 | 90 | 330-5300 |
| (TBA$^+$)$_x$FeSe | 2.1 | 240-500 | 60 | 48 | 320-6027 |

Most of the parameters in the table were collected by Koshelev *et al.*[3]. and Kacmarcik *et al*[4]. and Poole *et al*[5]. Gi quantifies the strength of the thermal fluctuations in superconductors. (TBA$^+$)$_x$FeSe is estimated to have one of the largest Ginzburg numbers among all iron-based superconductors. Notably, the Ginzburg numbers here are $(4\pi)^2$ times the Gi values in many previous works, and even some of the Gi values could have exceeded 1 here. The "unphysical" values do not have practical meanings but manifest the significant thermal fluctuations existing in those superconducting states. These modified Gi values are obtained via an equation that is corrected in SI units[6]:

$$Gi = (k_B T_c/\mu_0 H_c^2 \epsilon \xi^3)^2/2 = \frac{1}{8}(16\pi^2 \mu_0 \lambda^2 \gamma k_B T_c/\phi_0^2 \xi)^2$$

The penetration depth $\lambda_{ab}$ of (TBA$^+$)$_x$FeSe was determined to be 240-500 nm via the $\mu SR$ technique and low-$H$ magnetization measurements, and the anisotropy of superconductivity $\gamma = H_{ab}/H_c(T_c)$ was estimated to be greater than 60. Given $T_c = 48\ K$ and $\xi = 2.1\ nm$, the Gi value of (TBA$^+$)$_x$FeSe is calculated above 320.

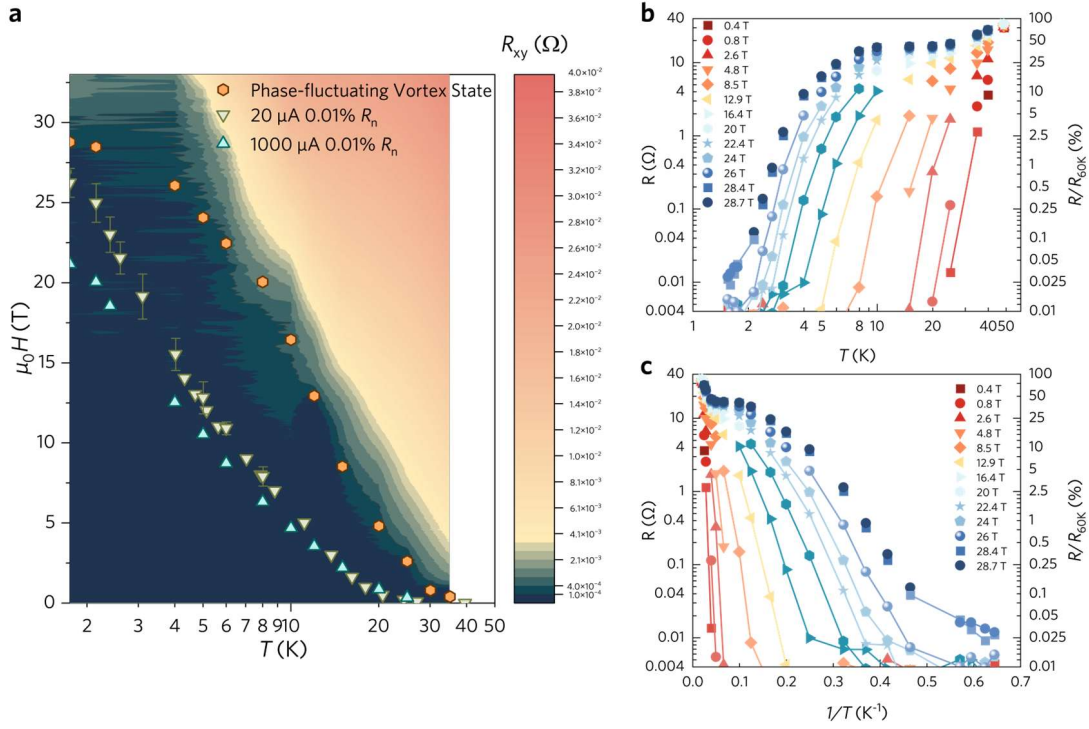

**Extended Fig. 2| The characteristics of the phase-fluctuating vortex states: finite resistance with vanishing Hall response. a,** The onset points of the phase-fluctuating vortex state and the irreversibility points were determined using 20 μA and 1000 μA excitation currents. The colored contour plot as a function of magnetic field and temperature is interpolated from Hall resistance ($R_{xy}$) measurements taken below 35 K. The onset points of the PFVS roughly outline the boundary at which the Hall resistance drops to the noise level, approximately at a magnitude of $10^{-4}$ $\Omega$. This generates a broad region with negligible Hall responses and substantial longitudinal resistances, as shown in the $H$-$T$ phase diagram. Notably, the Hall response seems to reemerge at higher magnetic fields above $\mu_0 H_{irr}(1.75\ K)$ at low temperatures. **b,** $R$-$T$ data points under several selected magnetic fields. **c,** $R$-$T^{-1}$ data points under several selected magnetic fields.

The points connected with lines represent the resistance with zero-Hall response within the phase-fluctuating vortex state regime. The other scatter points at higher temperatures are resistances with Hall response. Under relatively low magnetic fields, approximately 2.5–10% of the normal-state resistance $R_{60K}$ (1–5 Ω) has a particle–hole symmetry. Above 26 T, the ratio of the resistance with zero Hall response to the normal-state resistance is drastically suppressed, dropping from 5–10% to about 0.1%. The onset temperatures are also suppressed. Combined with the reemergence of the Hall resistance above 30 T at 1.75 K, as presented in Extended Fig.5a, it indicates the zero-Hall state is not the high-field ground state, different from the conventional AM state. Notably, the zero-Hall resistance does not exhibit a plateau on R-T and R-$T^{-1}$ plot, or neither has a power-law relation or a standard TAFF exponential temperature dependence, which means the PFVS doesn't originate from simple vortex dynamics.

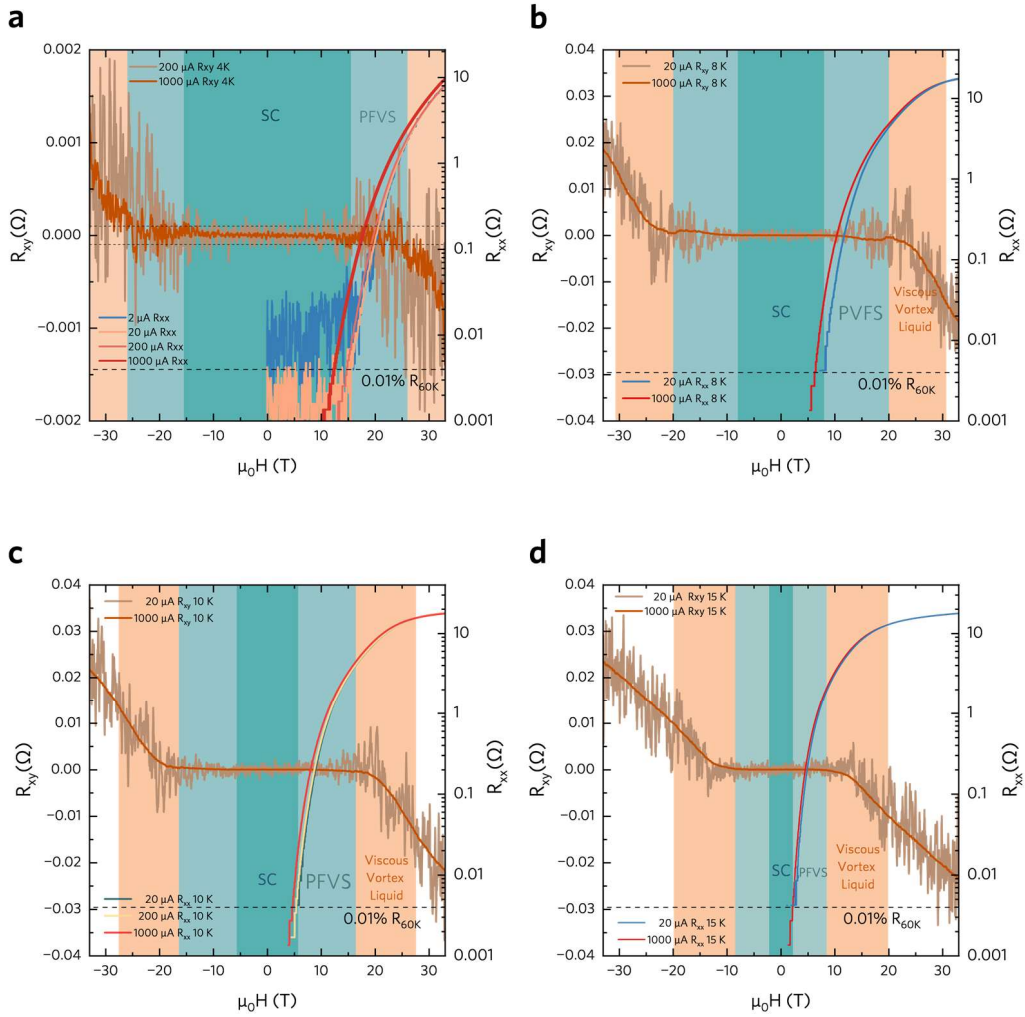

**Extended Fig. 3| Hall and longitudinal resistances measured with different excitation currents at 4 K, 8 K, 10 K and 15 K.** The dark cyan, light cyan and orange areas represent the superconducting state, phase-fluctuating vortex state (PFVS) and viscous (anomalous) vortex liquid state, respectively. The resistive values can be prominently influenced by different excitations beneath the critical magnetic field of the viscous vortex liquid. Given that the vortex motions could be easily tuned, we find that the onset magnetic fields of the PFVS remain nearly unaffected by external current excitations, despite the significant improvement in the signal–to–noise ratios (SNRs) of the Hall resistances with higher currents (1000 μA). This relative current-robustness of the Hall signals implies that the emergence of the PFVS may originate from strong vortex pinning in a intrinsically granular superconducting state, where the global vortices are frozen and the dissipation comes from occasionally vortex-like phase fluctuations.

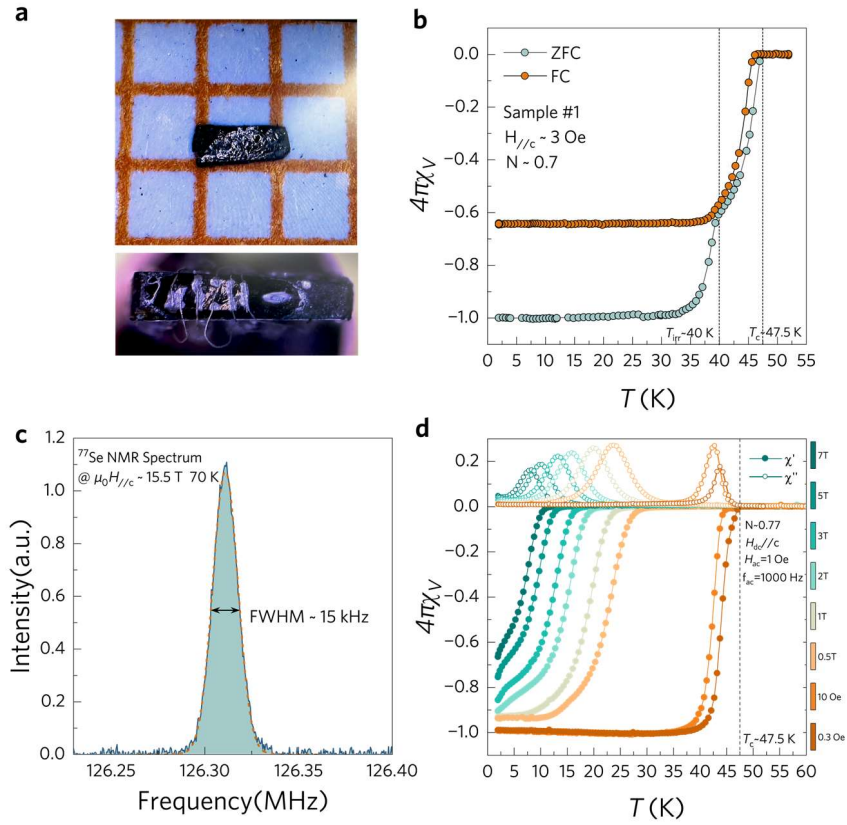

**Extended Fig. 4| Characterizations of high quality (TBA$^+$)$_x$FeSe samples. a,** Optical images of the cleaved (TBA$^+$)$_x$FeSe slice of sample #1 and the six-terminal electrode configuration. **b,** Field-cooled (FC) and zero-field-cooled (ZFC) DC magnetization curves of sample #1 measured at a 3 Oe out-of-plane magnetic field by a MPMS XL (Quantum Design). The onset temperature of the superconducting transition on magnetization is approximately 47.5 K, and the irreversible temperature is 40 K. A relatively small difference between the FC and ZFC curves implies the weak pinning of vortices in the sample, which also suggests that very few disorders are distributed on the superconducting planes. The magnetization is normalized to -1 by a demagnetization factor N ~ 0.7. **c,** NMR spectrum of $^{77}$Se nuclei measured at 15.5 T and 70 K, above the superconducting state. Its standard Gaussian line shape and narrow (15 kHz) full width at half maximum (FWHM) indicate a uniform local electronic environment for the $^{77}$Se nuclei, suggesting uniform chemical intercalation and electronic doping for the whole sample. **d,** AC magnetization curves of (TBA$^+$)$_x$FeSe at different magnetic fields. The real parts of the AC magnetization $4\pi\chi'$ at low magnetic fields were normalized to -1 with a demagnetization factor of N ~ 0.77. The superconducting transition temperature is determined at 47.5 K, which corresponds with the DC magnetization. The rapid jump in the real part $4\pi\chi'$ and the sharp peak in the imaginary part $4\pi\chi''$ confirm the uniformity of superconductivity at the penetration length scale at a near-zero magnetic field[1] and preclude large-scale chemical or structural inhomogeneities. The data in figures **c & d** were collected for (TBA$^+$)$_x$FeSe, which was intercalated from the same batch of pristine FeSe as sample #1.

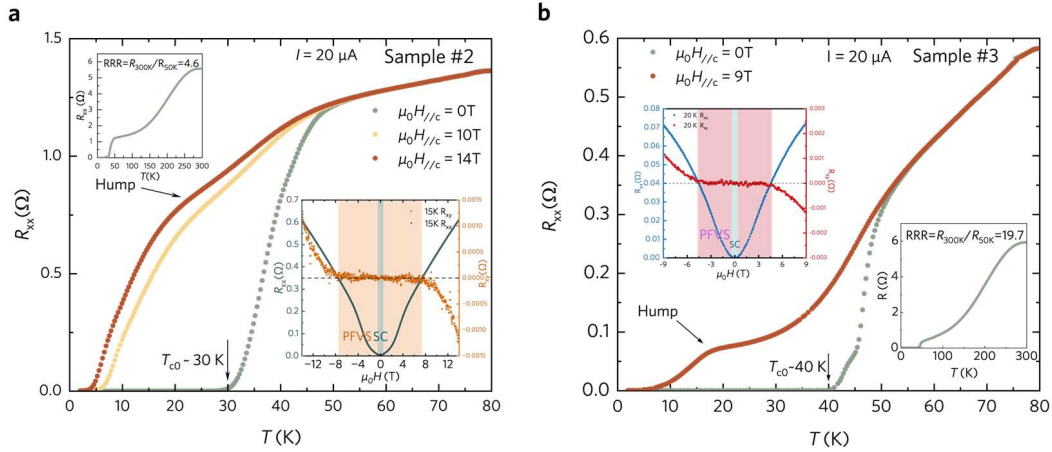

**Extended Fig. 5| Phase-Fluctuating vortex state behavior ($R_{xx} > 0$, $R_{xy} \sim 0$) discovered in different samples.** Our (TBA$^+$)$_x$FeSe samples are extremely air sensitive and can be aged if they are exposed to water or oxygen. The superconducting transition temperatures of various samples could be suppressed after aging. As shown in Fig. S9, Sample #2 has a relatively low coherent temperature of 30 K at zero magnetic field with a broader transition process, and Sample #3 has a two-step-like transition. However, the phase-fluctuating vortex state still manifests robustness in these samples with various $T_c$ and RRR values. The inserts show that when we measured the longitudinal and the Hall resistances simultaneously, the phase-fluctuating vortex states are between 0.6 T and 8 T at 15 K for Sample #2 and between 0.4 T and 4.6 T at 20 K for Sample #3. The humps on the resistance curves are also induced by magnetic fields.

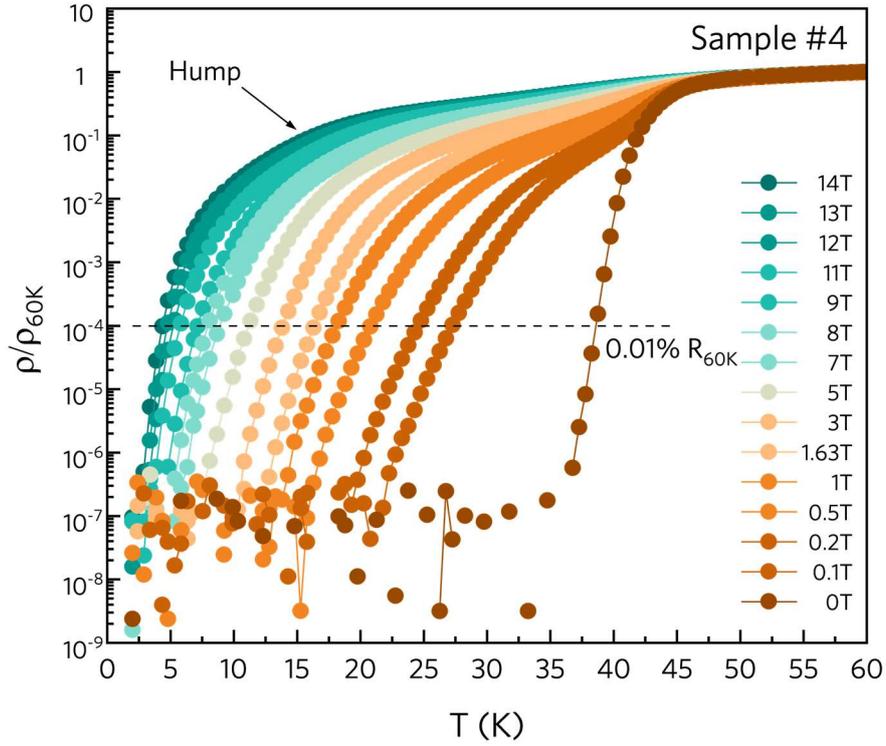

**Extended Fig. 6| R-T curves normalized by the resistivity at 60 K at magnetic fields up to 14 T.** Curves were measured for sample #4 with an excitation current of $I = 20\ \mu A$. Sample #4 was protected well from water and oxygen. This sample has a high superconducting fluctuation onset temperature of ~ 60 K and a high irreversibility point of ~ 39 K at near zero magnetic field, the same as that of Sample #1. The size and geometric profile of sample #4 are close to those of sample #1. The irreversibility points measured with $I = 20\ \mu A$ on the H-T phase diagrams in the text are combined with the data points collected on sample #1 in a water-cooled magnet and the data points on sample #4 in a DynaCool (Quantum Design). They are determined by the same criterion, 0.01% resistance at 60 K. The dashed line guides the eyes, representing the criterion of irreversibility points. The data at high magnetic fields connect with the data at low magnetic fields smoothly, which means that the experimental phenomena are reproducible in different $(TBA^+)_x FeSe$ samples.

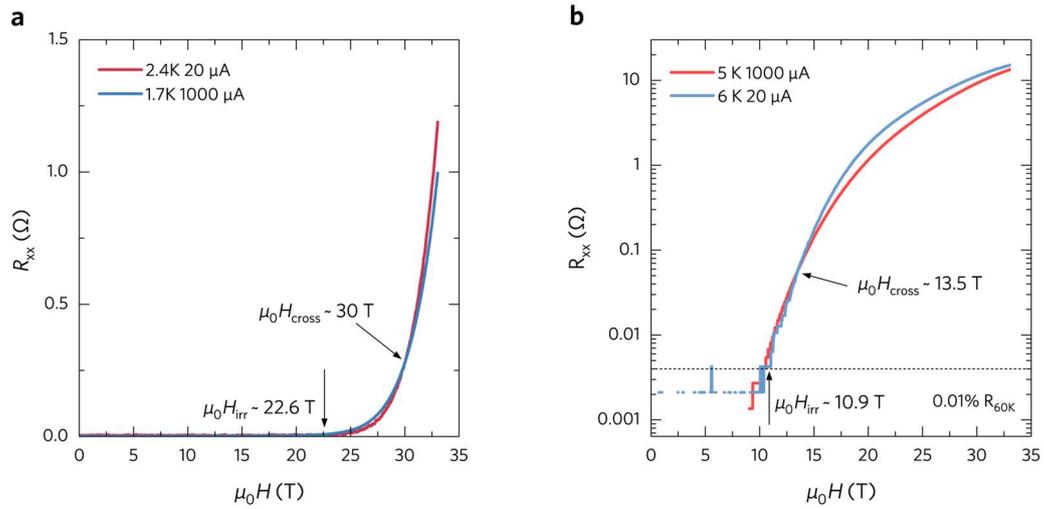

**Extended Data Fig. 7| The heating effect could be precluded by analysis of MR traces with the same irreversibility fields under various temperatures and excitation currents. a,** The magnetoresistance curves measured at 1.7 K ($I = 1000\ \mu A$) and 2.4 K ($I = 20\ \mu A$) have the same irreversibility field but only coincide at $\mu_0 H \approx 30\ T$. **b,** The magnetoresistance curves at 5 K and 6 K have the same irreversibility field but only coincide at $\mu_0 H \approx 13.5\ T$. It indicates that the suppression of irreversible fields by large currents at low temperatures is not due to heating effects. Detailed analysis is in Methods.

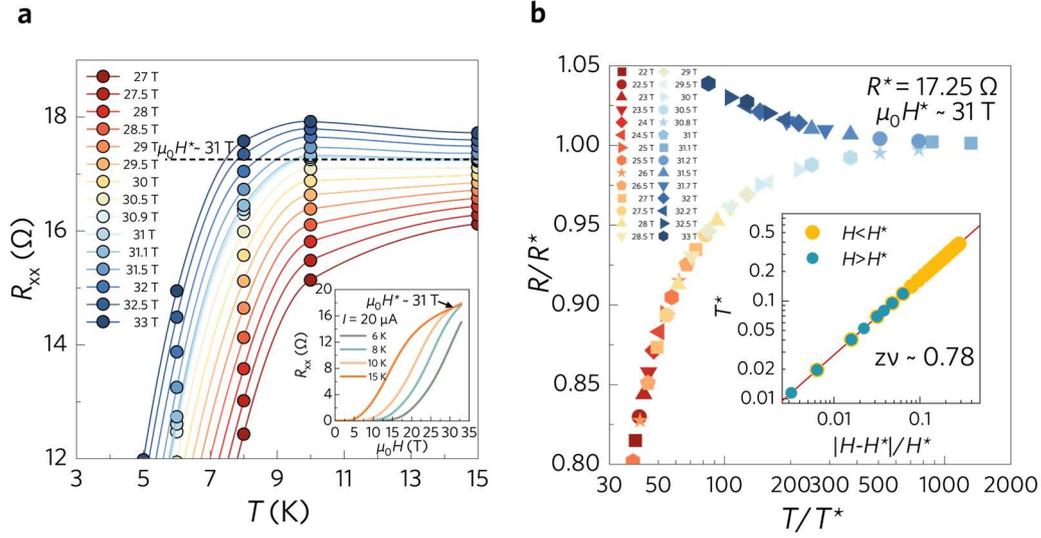

**Extended Fig. 8| The resistive humps on the R(T) curves indicate a quantum phase transition. a,** Evolution of the resistive hump at high magnetic fields. The inset shows the isothermal MR lines measured at 0 – 33 T with an excitation current $I = 20\ \mu A$. A crossing point of MR lines (10 K and 15 K) is presented at around 31 T. Notably, the crossing of MR lines measured at 10 K and 8 K is expected to be observed just above 33 T, the highest magnetic field of the magnet's capability. **b,** A finite-size scaling analysis for MR curves at 10 K and 15 K: $R_{xx}(H,T) = R^* f(T/T^*)$ ($R^* = 17.25\ \Omega$). The insert shows the scaling parameter $T^*$ as a function of $|H - H^*|/H^*$. It has a power law relation $T^* = |\delta|^{zv}$ with the reduced field parameter $\delta = \frac{|H-H^*|}{H^*}$. Here, we obtain a critical exponent $zv \sim 0.78$ ($z$ and $v$ are the dynamic and correlation length exponents, respectively) for this crossing point. This $zv$ value is close to the BG-VG transition in La$_{1.93}$Sr$_{0.07}$CuO$_4$[7], whose $zv = 0.737$, further implies the identity of a universal critical behavior: a 2D SIT in the clean limit governed by phase fluctuations. However, the whole quantum critical behavior cannot be fully clarified by this study. We only suggest that the resistive hump originates from the quantum phase transition of the superconductivity.